\documentclass[12pt,preprint]{aastex}

\begin{document}


\title{Dense, Fe-rich Ejecta in Supernova Remnants DEM L238 and DEM L249: A
New Class of Type Ia Supernova?}


\author{Kazimierz J. Borkowski}
\affil{Department of Physics, North Carolina State University,
    Raleigh, NC 27695}
\email{kborkow@ncsu.edu}
\author{Sean P. Hendrick}
\affil{Department of Physics, Millersville University, Millersville, PA 17551}
\author{Stephen P. Reynolds}
\affil{Department of Physics, North Carolina State University,
    Raleigh, NC 27695}


\begin{abstract}

We present observations of two LMC supernova
remnants (SNRs), DEM L238 and DEM L249, with 
the {\it Chandra} and {\it XMM-Newton} X-ray
satellites. Bright central emission, surrounded by a faint shell, is
present in both remnants. The central emission has an entirely thermal
spectrum dominated by strong Fe L-shell lines, with the deduced Fe
abundance in excess of solar and not consistent with the LMC
abundance. This Fe overabundance 
leads to the conclusion that DEM L238 and DEM L249 are remnants
of thermonuclear (Type Ia) explosions. The shell emission originates in 
gas swept up and
heated by the blast wave. A standard
Sedov analysis implies about 50 $M_\odot$ in both swept-up shells,
SNR ages between 10,000 and 15,000 yr, low ($\lesssim 0.05$ cm$^{-3}$)
preshock densities, and subluminous explosions with energies of $3
\times 10^{50}$ ergs. The central Fe-rich supernova ejecta are close
to collisional ionization equilibrium. Their presence is unexpected, because
standard Type Ia SNR models predict faint ejecta emission with short
ionization ages. Both SNRs belong to a previously
unrecognized class of Type Ia SNRs characterized by bright interior
emission.  Denser than expected ejecta and/or a dense circumstellar
medium around the progenitors are required to explain the presence of
Fe-rich ejecta in these SNRs. Substantial amounts of circumstellar gas
are more likely to be present in explosions of more massive Type Ia
progenitors. DEM L238, DEM L249, and similar SNRs could be remnants of
``prompt'' Type Ia explosions with young ($\sim 100$ Myr old) progenitors.

\end{abstract}


\section{Introduction}

Heavy elements are produced in stars and then injected into the
interstellar medium (ISM) by the combined action of stellar winds and
stellar explosions.  Thermonuclear and core-collapse explosions are
dominant sources of heavy elements, and they govern the evolution of
chemical abundances in galaxies. Thermonuclear (Type Ia) explosions
inject iron-rich supernova (SN) ejecta into the ISM, while ejecta of
core-collapse (CC) SNe contain large quantities of oxygen and other
$\alpha$-elements. Heavy elements are dispersed by remnants of SNe and
eventually mix with the ambient ISM, forming subsequent generations of
stars. The O/Fe (or more generally $\alpha$-elements/Fe) ratio depends
on the relative frequency of CC and Type Ia explosions, and on how a
galaxy has evolved in the past.  It is now well established that this
ratio varies among galaxies.  In particular, the $\alpha$/Fe ratio is
lower in the Magellanic Clouds (MCs) than in our Galaxy
\citep[e.g.,][]{pompeia06}. This suggests that chemical enrichment by
Type Ia SNe has been relatively more important in the MCs than in our
Galaxy. It is not clear whether this is caused by an increased
frequency of Type Ia versus CC explosions in the MCs \citep{tsijimoto95}
or by differences in evolutionary histories between the MCs and our
Galaxy \citep{pata98}. Our poor understanding of progenitors of Type
Ia explosions is a significant problem as well; the recent proposal of
a new class of ``prompt'' Type Ia SNe \citep*{mdp06,sullivan06} associated with
relatively young ($\sim 100$ Myr) stars may force reevaluation of
current ideas about the poorly understood chemical evolution of the
MCs.

Studies of remnants of Type Ia explosions in the MCs allow us to
observe injection of Fe into the ISM directly and to study SN ejecta in
detail, leading to a better understanding of Type Ia explosions and
the enrichment of the MCs in heavy elements.  \citet{tuohy82}
discovered four supernova remnants (SNRs) in the Large Magellanic Cloud
(LMC) with strong Balmer lines of hydrogen and weak forbidden lines of
heavy elements; such Balmer-dominated remnants are believed to be of
Type Ia origin.  X-ray observations with the {\it Advanced Satellite
for Cosmology and Astrophysics (ASCA)} confirmed a Type Ia origin for
Balmer-dominated SNRs 0509-67.5 and 0519-69.0 by detecting strong Fe
lines in their X-ray spectra \citep{hughes95}. Their ages are 400 
and 600 yr, respectively, based on the recent discovery of light echoes
from these explosions \citep{rest05}. The other two Balmer-dominated
SNRs, DEM L71 and 0548-70.4, were also found to be rich in Fe
\citep{hughes98,hendrick03}.  There are several other Type Ia SNRs in
the MCs, as indicated by their Fe overabundance.  The bright SNR N103B is
only $\sim 1000$ yr old, based on one detected light echo
\citep{rest05}; its origin is still under debate, with
\citet{vanderheyden02} and \citet{lewis03} advocating a CC and Type
Ia origin, respectively.

Most progenitors of Type Ia SNe are believed to belong to old stellar
populations.  Because at most a modest amount of circumstellar
material may be associated with the pre-explosion evolution of their
low-mass progenitors, SN ejecta dissipate their kinetic energy mostly
interacting with the ambient ISM.  Lack of a dense circumstellar
medium for most Type Ia SNe is consistent with the absence of radio
emission from SNe Ia \citep{weiler02}.  Type Ia SN progenitors are
usually located far from clumpy star formation regions and generally
explode in a relatively uniform diffuse ISM.  Remnants of historical
Type Ia SNe of Tycho and 1006 indeed exhibit rather smooth outer blast
waves. \citet{badenes06} successfully modeled the Tycho X-ray spectrum
with SN ejecta colliding with an ambient medium of constant density,
finding good agreement for delayed-detonation models of Type Ia
explosions. This encouraging agreement gives credence to the
conventional wisdom in which most Type Ia SNRs arise from the
interaction of SN ejecta with a rather uniform ambient medium, without
needing to invoke the presence of dense circumstellar material in the
immediate vicinity of their progenitors. Hydrodynamic models for
such an interaction have been considered in the past
\citep[e.g.,][]{dwarkadas98}, and an extensive grid of models
(including synthetic X-ray spectra) has become recently available
\citep*{badenes03,badenes05,badenes06}.

We present {\it Chandra} and {\it XMM-Newton} observations of two
moderately ($\sim 10^4$ yr) old SNRs in the LMC, DEM L238
(\objectname{B0534-70.5}) and DEM L249 (\objectname{B0536-70.6}),
which unexpectedly reveal the presence of Fe in their interiors,
presumably produced by thermonuclear explosions of their progenitors.
These $3\arcmin$--diameter SNRs have been previously studied at
optical and X-ray wavelengths \citep{mathewson83,williams99}, but
these studies have not revealed anything unusual, except for the X-ray
emission enhancement in their interiors. We find that the presence of
Fe-rich ejecta at their centers is not consistent with the standard Type
Ia SNR models just mentioned. The presence of denser than expected
material in the vicinity of their SN progenitors or in SN ejecta is
required. We discuss several possible scenarios that could explain
this finding.

\section{Observations and Imaging}

\begin{figure}
\epsscale{1.0}
\plotone{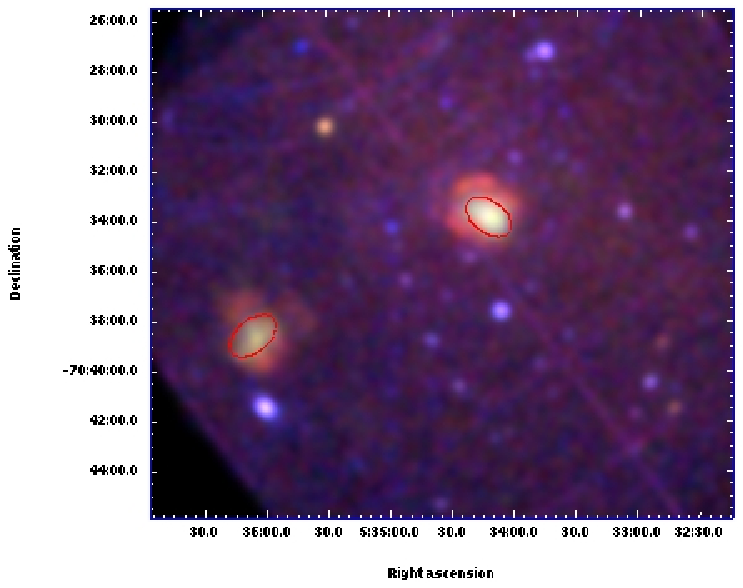}
\caption{
{\it XMM-Newton} EPIC (pn+MOS1+MOS2) image of the field of view 
encompassing SNRs DEM
L238 ({\it center}) and DEM L249 ({\it to the east}). Data have been
convolved with a $23\farcs5$ FWHM Gaussian.  Soft (0.3--0.7 keV),
medium (0.7--1.2 keV), and hard (1.2--7 keV) photons are coded by red,
green, and blue, respectively. Pronounced spectral differences between
prominent central emission and outer shells are apparent for both
SNRs.  Ellipses indicate central spectral extraction regions.
\label{fig1}}
\end{figure}

X-ray observations with {\it XMM-Newton} were done on 2003 July 23, 
with the primary target DEM L238 located at the center of the 
European Photon Imaging Camera (EPIC) MOS
and pn cameras. The total exposure time was 43.5 ks, and the Medium
filter was used. We used the {\it XMM-Newton} Science Analysis System
(SAS) software, version 6.5.0 to screen data for periods of high flare
events, extracted images and spectra, and generated primary and
secondary spectral response files. After screening of the data for
flares, the effective exposure time was 17.4 ks for the pn camera,
and 21.8 ks each for the MOS1 and MOS2 cameras. We combined EPIC MOS
and pn events to create an X-ray image (Fig.~\ref{fig1}), not
corrected for variations in the effective exposure time across the
EPIC field of view. Linear features seen in this image are artifacts 
at boundaries between individual CCD chips comprising pn and MOS 
cameras.  

In addition to our primary target DEM L238, SNR DEM L249 is also
located in the EPIC field of view (Fig.~\ref{fig1}). Both SNRs exhibit
shells of soft X-ray emission surrounding central emission
enhancements with harder X-ray spectra. (The shell morphologies seen in
Fig.~\ref{fig1} have been slightly distorted by linear artifacts 
located at CCD boundaries crossing both shells.) Patchy emission seen in
optical images \citep{mathewson83} coincides with the soft 
X-ray-emitting shells; this is the swept-up ISM. Optical emission presumably
traces localized dense regions where the blast wave slowed down and
became radiative; the soft X-ray-emitting shell reveals the bulk of
the swept-up ISM gas heated to X-ray-emitting temperatures by the
primary blast wave. 

{\it Chandra} observed DEM L238 for 59.2 ks on 2003 October 25--26, 
with the ACIS S3 chip. DEM L249 was in the field of view of the ACIS
I2 and I3 chips, so high spatial resolution images are available for
both SNRs. We used the {\it Chandra} X-ray analysis software CIAO,
version 3.3 to screen data for periods of high background, extracted images
and spectra, and generated spectral response files. Very faint mode
was used to reduce the particle background. After screening, the
effective exposure time was 50.4 ks for DEM L238 and 47.5 ks for
DEM L249. {\it Chandra} images of DEM L238 and DEM L249 are shown in
Figures \ref{fig2} and \ref{fig2a}. No correction has been applied to 
account for variations of the effective exposure time across the ACIS chips.

The high spatial resolution {\it Chandra} image of DEM L238 shows the
separation between the soft X-ray-emitting shell and the harder
interior emission more clearly. The interior emission is elliptical in
shape, offset from the center of the shell toward the southwest.  The shell and
the central emission partially overlap. The soft X-ray-emitting shell
of DEM L249 is faint, so only its brightest sections are seen in the
{\it Chandra} image. The boundary between the I2 and I3 chips runs
diagonally (southeast-northwest) up across the image, between the northeast shell and the central
emission; this produces a linear artifact in the remnant's X-ray
image.

A montage of high-quality narrowband optical images obtained in the
Magellanic Cloud Emission Line Survey\footnote{Preliminary data sets
and description of this survey can be found online at
http://www.ctio.noao.edu/mcels.}  \citep[MCELS;][]{smith00} is shown
for each SNR in Figures \ref{fig2} and \ref{fig2a}. Optical emission
and soft X-rays in DEM L238 have similar shell-like morphologies with
significant departures from a spherical shape, but they do not
correlate well at small spatial scales. This is expected as optical
emission is produced in slow radiative shocks and traces the location
of dense ISM, while X-rays are emitted by gas heated to high
temperatures in much faster nonradiative shocks. No optical emission
is associated with the interior X-ray emission in DEM L238
(Fig.~\ref{fig2}).  DEM L249 is much more asymmetric in the optical,
elongated in the southwest-northeast direction. Optical filaments are present at
its center, but there is only a weak correlation between them and the
central X-ray emission. The soft X-ray-emitting gas in the northeast and southwest
(Figs.~\ref{fig1} and~\ref{fig2a}) again matches well with the
optically emitting gas on large spatial scales, but poorly on small
scales. Somewhat faster radiative shocks in the northwest than in the southeast are
implied by the predominance of [\ion{O}{3}] $\lambda 5007$ emission in the northwest,
and its weakness relative to [\ion{S}{2}] $\lambda\lambda 6716, 6731$ and
H$\alpha$ emission in the southeast. A large-scale northwest-southeast density gradient in the
ISM is likely responsible for these differences in the radiative shock
speeds. Soft X-ray emission farther northwest, seen only in more sensitive
{\it XMM-Newton} observations (Fig.~\ref{fig1}), has no visible optical
counterpart, presumably because the ISM density there is too low for
the formation of radiative shocks.

\section{Spectral analysis}

We extracted central and shell spectra from {\it XMM-Newton} and {\it
Chandra} data in order to deduce the properties of the blast waves and
establish the nature of the interior emission in DEM L238 and DEM
L249. The central extraction regions are shown in Figures \ref{fig1}
and \ref{fig2}. The {\it Chandra} shell region of DEM L238 encompasses
the whole remnant minus its central region.
Background spectra were extracted from adjacent,
point-source-free regions of the same CCD chips.  Spectral analysis
was performed with XSPEC, version 12 \citep{arnaud96}. We used both the
nonequilibrium ionization (NEI), version 2.0 thermal models, based on the
astrophysical plasma emission code and database
\citep[APEC/APED;][]{smith01} and augmented by addition of
inner-shell processes \citep{badenes06}, and the collisional
ionization equilibrium APEC models. Spectra were binned to a minimum
25 counts bin$^{-1}$ in order to allow the use of $\chi^2$ statistics.

\subsection{Shell Emission}

The interior and shell {\it Chandra} spectra of DEM L238 have 6676 and
3772 counts in the energy range 0.4--3.0 keV, or 6120 and 2380 counts
after background subtraction. They differ greatly
(Fig.~\ref{fig3}), as expected from the pronounced color differences seen
in Figures \ref{fig1} and \ref{fig2}. The presence of emission lines
in these spectra shows that the observed X-ray emission is of thermal
origin. We fit the shell spectrum with a standard Sedov model with 0.4
solar (cosmic) abundances appropriate for the LMC. We used a
two-component absorption model: foreground Galactic absorption with the
column density of $N_H = 6 \times 10^{20}$ cm$^{-2}$ appropriate for
this location on the sky \citep{stav03}, and the LMC absorption with
0.4 solar abundances that we allowed to vary. The best fit is shown
in Figure \ref{fig3}, with postshock mean temperature $kT_s =
0.40$ keV [$(0.34, 0.53)$, 90\%\ confidence interval], postshock
electron temperature $kT_e = 0.11 (0.0, 0.34)$ keV, ionization age
$\tau = n_e t_{SNR} = 9.4 (7.1, 14) \times 10^{11}$ cm$^{-3}$ s (where
$n_e$ is the postshock electron density and $t_{SNR}$ is the SNR age),
and emission measure $EM = \int n_e n_H dV$ of $4.4 \times
10^{57}$ cm$^{-3}$ (at the assumed 50 kpc distance to the LMC). The
shock speed $V_s$ corresponding to a temperature of 0.40 keV is 600 km
s$^{-1}$.  No internal absorption within the LMC was necessary. While
the Sedov fit with standard LMC abundances was statistically
acceptable, an inspection of Figure \ref{fig3} reveals that the Sedov
model underpredicts the observed spectrum at low energies, and the
K$\alpha$/Ly$\alpha$ line ratio of O is underestimated. Apparently
there is a substantial amount of low-temperature plasma in the blast
wave of DEM L238, in excess of what is predicted by a standard Sedov
model. This is likely caused by variations in the blast wave speed
around the remnant's periphery, with the very soft X-ray emission
coming from shocks with speeds less that the average blast wave speed
of 600 km s$^{-1}$, but more than $\sim 100$ km s$^{-1}$ expected for
radiative shocks seen in optical.

In the Sedov dynamical model, $V_s = 2R_s/5t_{SNR}$, where $R_s =
80\arcsec$ (20 pc at 50 kpc distance) is the remnant's radius. With
the current shock speed of 600 km s$^{-1}$, the estimated remnant's
age $t_{SNR}$ is equal to 13,500 yr. The preshock density $n_0$ ahead
of the blast wave in DEM L238 can be estimated from the Sedov EM
\citep[e.g.,][]{ham83}. Our Sedov spectral fit to the shell spectrum
of DEM L238 (Fig.~\ref{fig3}) excluded the central region shown in
Figure~\ref{fig2}, so that we need to correct the EM obtained in this
fit to account for the incomplete spatial coverage of the remnant. An
X-ray-emitting shell in a Sedov model may be approximated as a
geometrically thin shell, and the correction factor is equal to
$1/(1-2f_c)$, where $f_c$ is the fraction of the solid angle occupied
by the central region as seen from the remnant's center, and the
factor of 2 accounts for the front and back sections of the shell (we
assumed a spherically symmetric SNR).  After 60\%\ correction to the
EM, we arrive at $n_0 = 0.05$ cm$^{-3}$. The estimated swept-up mass
is 60 $M_\odot$.  The inferred kinetic energy $E_k$ of the explosion
is equal to $3.3 \times 10^{50}$ ergs. This is less than the canonical
explosion energy of $10^{51}$ ergs, but in agreement with energies
derived for several Type Ia SNRs in the LMC
\citep{hendrick03,ghavamian03}. The product of $n_0$ and $t_{SNR}$ is
equal to $2 \times 10^{10}$ cm$^{-3}$ s, about 40 times less than the
Sedov ionization age derived directly from the spectral fit. This
suggests longer $t_{SNR}$ and/or higher $n_0$ than estimated
above. Dense ISM is present near DEM L238, as evidenced by large-scale
radiative shocks seen in the optical (Fig.~\ref{fig2}), and the X-ray
emission might be produced by gas with densities higher than 2
cm$^{-3}$ (this requires a lower volume filling factor than the 1/4
appropriate for the standard Sedov model).  These inconsistencies of
the Sedov model cast some doubt on its validity for DEM L238. A
crucial assumption in this model is the uniformity of the ambient
medium around the SN progenitor; this assumption might be appropriate
for most Type Ia explosions but would be incorrect for a massive
progenitor of a CC SN. A CC progenitor is expected to modify the
ambient medium through the action of its stellar winds prior to the SN
explosion. In this case, inferences based on the Sedov analysis could
be incorrect.

The X-ray-emitting shell in DEM L249 is too faint for spectral
analysis with {\it Chandra}, so we used {\it XMM-Newton} data for this
purpose. In Figure \ref{fig4} we contrast {\it XMM-Newton} EPIC pn
spectra of the shell and the interior emission. The spatially integrated 
{\it Chandra} spectrum is also shown. This {\it Chandra} spectrum does not
completely include the faint shell sections in the northeast and northwest, and it is 
dominated by the bright interior emission. 
The difference between the central and shell
spectra is again obvious.  We extracted the shell spectra from the northeast
and south part of the shell, not from the fainter northwest section seen in 
Figure~\ref{fig1}, where the SNR is not apparent on the MCELS optical images 
(Fig.~\ref{fig2a}) and where
the blast speed might be higher than in the northeast and south. The total
numbers of counts in the 0.4--3.0 keV range were 815, 321, and 309 for
pn, MOS1, and MOS2 detectors, respectively.  After background subtraction, these
numbers were reduced to about 600 counts for pn, and $\sim 225$ each
for MOS detectors.  Because of the low numbers of counts, we made
joint fits to pn, MOS1, and MOS2 spectra with the Sedov model (only pn
spectra are shown in Fig.~\ref{fig4}).  Foreground Galactic absorption
with $N_H = 6 \times 10^{20}$ cm$^{-2}$ was assumed, based on \ion{H}{1}
observations by \citet{stav03}. A Sedov model fit with 0.4 solar
abundances was not acceptable, primarily because this model could not
reproduce Fe emission lines at $\sim 0.8$ keV seen in the pn shell
spectrum of Figure \ref{fig4}. A better fit resulted by allowing the
Fe abundance to vary, with the Ni abundance tied to Fe, although the
lack of internal consistency between pn, MOS1, and MOS2 spectra
prevented a statistically acceptable fit. This fit is shown in Figure
\ref{fig4}. Again, no internal absorption in the LMC was required. The
derived postshock temperature is 0.36 keV, corresponding to 550 km
s$^{-1}$, and the ionization timescale $\tau = 2.1 \times 10^{11}$
cm$^{-3}$ s.  The Fe abundance is 1.5 solar, significantly in
excess of the LMC Fe abundance. The inferred Fe overabundance in the
swept-up ISM might not be real. DEM L249 has relatively bright
central emission with strong Fe L-shell emission (see
Fig.~\ref{fig4}) and a much fainter shell, and its far off-axis
location, where the {\it XMM-Newton} point-spread function is quite
broad, increases the potential for contamination of the shell spectrum
by the central emission. In view of the rather poor photon statistics,
we have not attempted to deconvolve the central and shell spectra. The
derived shock speed in DEM L249 is only slightly lower than in DEM
L238, while the ionization age is shorter by a factor of 4. These
similarities, as well as similar central spectra (cf.~Figs.~\ref{fig3} 
and \ref{fig4}), suggest these two SNRs are alike in
many respects.

\subsection{Central Emission}

Central X-ray spectra of DEM L238 extracted from the {\it XMM-Newton}
EPIC pn and combined MOS detectors and from the {\it Chandra} ACIS S3
chip are shown in Figure \ref{fig5}. (We present an average of MOS1
and MOS2 spectra for display purposes only; all fits were done using
separate spectral responses for MOS1 and MOS2 detectors.) The pn, MOS, 
and ACIS S3 spectra differ in shape because of variations among 
individual CCD spectral responses. As can be
seen in the high spatial resolution {\it Chandra} image of DEM L238
(Fig.~\ref{fig2}, {\it left}), the soft X-ray-emitting shell and the
harder interior emission overlap spatially. The central spectra are
then comprised of a sum of the shell and interior emission, so in
addition to a single {\tt vpshock} model \citep{borkowski01} we also
used a two-component model in spectral fitting. For the shell
component, we fixed parameter values to those found in the fits to the
shell spectrum described above, allowing only the normalization of
this component to vary. For the second interior component, we
initially used various NEI models, but the long ionization timescales
derived from two-component fits always suggested that gas at the
center is close to collisional ionization equilibrium (CIE). We found
that use of the CIE APEC thermal model produced as good statistical
fits as NEI thermal models, so we present here fits with the APEC
model (plus the Sedov shell component). We assumed the same
interstellar absorption, $N_H = 6 \times 10^{20}$ cm$^{-2}$, as in
fits to the shell spectrum. We made separate fits to {\it Chandra}
ACIS S3, {\it XMM-Newton} EPIC pn spectra, and joint fits to {\it
XMM-Newton} EPIC MOS1 and MOS2 spectra.

All fits with the APEC model with 0.4 solar abundances appropriate for
the LMC failed to produce acceptable results. Large residuals seen
near 0.9 keV suggest the presence of stronger than expected Fe
L-shell lines. We obtained statistically acceptable fits by allowing
the Fe abundance to vary, with the Ni abundance tied to Fe.
These fits are presented in Table~\ref{tabl238} and Figure
\ref{fig5}. The Fe L-shell line complex at $\sim 0.9$ keV is prominent
in all three spectra. The X-ray-emitting gas at the center of DEM L238
is rich in Fe. This appears to be a robust result, as we arrived at a
similar Fe overabundance when we allowed for either underionized ({\tt
vpshock} and {\tt vnei} NEI models in XSPEC) or overionized
(custom-made XSPEC models) plasmas. The most straightforward
interpretation of this result is that we are seeing Fe-rich SN ejecta,
presumably produced in a thermonuclear (Type Ia) explosion.

The relative abundances of elements in SN ejecta are of great
interest, but their determination is difficult in DEM L238 because of
the spatial overlap of the interior and the shell emission and
uncertainties in atomic data for Fe L-shell lines. For the models
shown in Figure \ref{fig5} and listed in Table \ref{tabl238}, the O/Fe
ratio with respect to the solar value of \citet{andgr89}, [O/Fe] is
0.29, 0.31, and 0.26 in fits to {\it XMM-Newton} MOS1+MOS2, pn, and
{\it Chandra} spectra, respectively.  This gives the O/Fe ratio of 5.3, 5.7, 
and 4.7 by number, higher than found in most Type Ia SN models
\citep[e.g., the O/Fe ratio ranges from 0.25 to 0.8 in models of][]{iwa99}, 
but far lower than the expected value of order
70 for CC events.  However, we find that even very low O/Fe ratios
are consistent with observations, based on additional fits (not shown)
where we allowed the O abundance within ejecta to vary. In these fits
no O was required; however, errors were large.  In our two-component
model fits, the shell component with the prominent O lines dominates
at low energies, while the interior ejecta component is very strong in
the Fe L-shell lines. The shell contribution is not accurately
known, so determination of the O/Fe ratio within the ejecta is coupled
to the problem of separation of the shell and ejecta
contributions. This results in relatively large statistical errors on
the derived O/Fe ratio.

A reliable determination of elemental abundances in the Fe-rich gas of
DEM L238 is also made difficult by continuing uncertainties in atomic
data for Fe L-shell lines. This problem is particularly severe for
elements such as Ne and Mg, whose X-ray lines overlap with the Fe lines
at the low spectral resolution of the CCD detectors. We made
additional fits (not shown), in which we allowed Ne and Mg abundances to
vary. In these fits, no Ne was required, while the Mg abundance was
approximately solar. These two-component fits with variable Ne and Mg
abundances were of much better quality than the two-component fits
presented in Table \ref{tabl238}, as judged by their lower $\chi^2$
values.  The low Ne abundance could be real, consistent with low Ne
abundances in Type Ia SN models.  The relatively high Mg abundance
might be spurious, because a number of Fe lines spectrally overlapping
with Mg lines are missing in APEC and other existing X-ray spectral
codes \citep{brickhouse00}.  Such deficiencies in Fe L-shell atomic
data pose serious problems for studies of the ejecta composition in
DEM L238.

It is also difficult to determine whether we observe pure
heavy-element ejecta, or whether substantial mixing occurred with
gas with normal LMC abundances, where most electrons are contributed by
H and He. Mixing was implicitly assumed when we used the APEC model
with abundances of elements other than Fe and Ni fixed to 0.4 solar in
fits to the central emission. In this case, about $\sim 40 f^{1/2}$
M$_\odot$ (including $\sim 0.1 f^{1/2}$ M$_\odot$ of Fe) would be
present in the central region ($f$ denotes the unknown filling
fraction of the X-ray-emitting gas). This mass estimate should be
considered as an upper limit, because errors in the determination of
the Fe abundance are large. In particular, we believe that the true
upper limits on the Fe abundance are much larger than the statistical
upper limits listed in Table \ref{tabl238}. Our spectral model for the
Fe L-shell emission is incomplete, and we expect this incompleteness
to become increasingly important with the increasing Fe abundance. The
formal upper limits on the Fe abundance derived from the $\chi^2$
analysis may then be highly misleading. We do not exclude the
possibility that ejecta consist of pure Fe, in which case as much as
$\sim 1$ M$_\odot$ of Fe could be present at the center of DEM
L238. The most likely intermediate case involves several solar masses
of gas, including a few tenths M$_\odot$ of Fe.

Central X-ray spectra of DEM L249 extracted from the {\it XMM-Newton}
EPIC pn and combined MOS detectors and the spatially integrated {\it
Chandra} spectrum are shown in Figure \ref{fig6}. There are 2290
(2130), 1526 (1400), and 3171 (2180) counts in these spectra before
(after) the background subtraction. We used the same techniques in
analyzing these spectra as for DEM L238. Results of spectral fits are
presented in Table \ref{tabl249} and Figure \ref{fig6}. These spectra
are dominated by Fe L-shell emission, and again enhanced Fe
abundances are required. Determination of reliable O, Ne, and Mg
abundances with respect to Fe within the ejecta faces the same
obstacles as in DEM L238, and the derived abundance ratios are the
same as in DEM L238 (albeit within large errors).  The gas temperature
is slightly lower than in DEM L238, producing a noticeable shift in
the centroid of the Fe L-shell emission toward lower energies when
compared with DEM L238. DEM L249 appears to be very similar to DEM
L238, but at a slightly more advanced evolutionary stage, as indicated
by its larger angular size, slower shock speed, cooler Fe-rich gas
at its center, and lower X-ray flux.
 
\section{Discussion}

Our main finding is the discovery of Fe-rich gas in the centers of the 
moderately old ($\sim 10^4$ yr) SNRs DEM L238 and DEM L249. An
important property of this gas is its very long ionization age; we
found the interior X-ray spectra of both SNRs well fit by collisional
equilibrium ionization 
models. Such remnants, with central Fe overabundances and long
ionization timescales, have been found before. In addition to four
Balmer-dominated SNRs and N103B mentioned in \S~1, several
other Fe-rich remnants have been discovered in the MCs: 0534-69.9
\citep{hendrick03}; DEM S128, IKT5, and IKT25 \citep{vander}; DEM L316A
\citep{nishiuchi01,williams05}; and 0454-67.2 \citep{seward06}. Among
them, DEM L316A is most similar to DEM L238 and DEM L249, as the
Fe-rich gas at its center is in ionization equilibrium, and a swept-up
shell is present. The SMC SNRs DEM S128 and IKT 5 also have Fe-rich
gas in CIE, but their shells have not been detected by {\it
XMM-Newton}.  There is then a whole class of remnants with
centrally located Fe-rich gas in CIE, and faint or undetectable X-ray
shells, making up a substantial fraction of all MC Type Ia SNRs.

It is usually assumed that the presence of Fe-rich ejecta is a
signature of a Type Ia explosion.  However, DEM L238 and DEM L249
exhibit a number of puzzling properties that are hard to understand in
the framework of the standard Type Ia models mentioned in \S~1.
The long ionization times of the Fe-rich gas imply that
the gas density has had to be high enough in the past. In the standard
Type Ia models of \citet{badenes03,badenes05,badenes06}, this can be 
accomplished only by
increasing the ambient medium density (assumed to be constant). But
the predicted shell emission in such models is orders of magnitude
brighter than observed. If we use instead the blast wave properties
derived from the Sedov fits to the shell emission, which imply low ISM
densities, then the Fe-rich ejecta are always at low densities and
ionization timescales are short. We demonstrate this by considering
simple models with an exponential density profile and uniform
ambient ISM \citep{dwarkadas98}, with ejecta mass of 1.4 M$_\odot$
composed of pure Fe, and with $E_k = 3.3 \times 10^{50}$ ergs, $n_0 =
0.05$ cm$^{-3}$, and $t_{SNR} = 13,500$ yr derived from the Sedov
analysis of DEM L238. We performed our one-dimensional simulations 
with the Virginia Hydrodynamics PPM (VH-1) code. This code was written 
and tested by the Virginia Numerical Astrophysics Group and is based 
on the Lagrangian-Remap version of the piecewise parabolic method 
developed by \citet{colwoo84}. We employed a simplified method devised
for multidimensional hydrodynamic simulations coupled with NEI
calculations to follow Fe ionization states within the ejecta. We used
the NEI, version 1.1 spectral code in XSPEC to calculate X-ray spectra
\citep[Fe L-shell atomic data in all NEI models in XSPEC are based on
theoretical calculations of][]{liedahl}.  (The low-ionization timescales
we found led us to use NEI, version 1.1 instead of the APEC-based version 
2.0, because
the former includes more of the lower ionization transitions.)
The calculated X-ray spectra
are compared with the central {\it Chandra} X-ray spectrum of DEM L238
in Figure \ref{fig7}. The failure of a standard Type Ia model is
obvious, as the shocked Fe in the model is underionized and the ejecta
X-ray emission is more than an order of magnitude too faint. An
inspection of more realistic Type Ia models of 
\citet{badenes03,badenes05,badenes06}
confirms this conclusion. The fundamental problem is that the density
of the Fe-rich ejecta is too low.

\begin{figure}
\epsscale{1.0}
\plotone{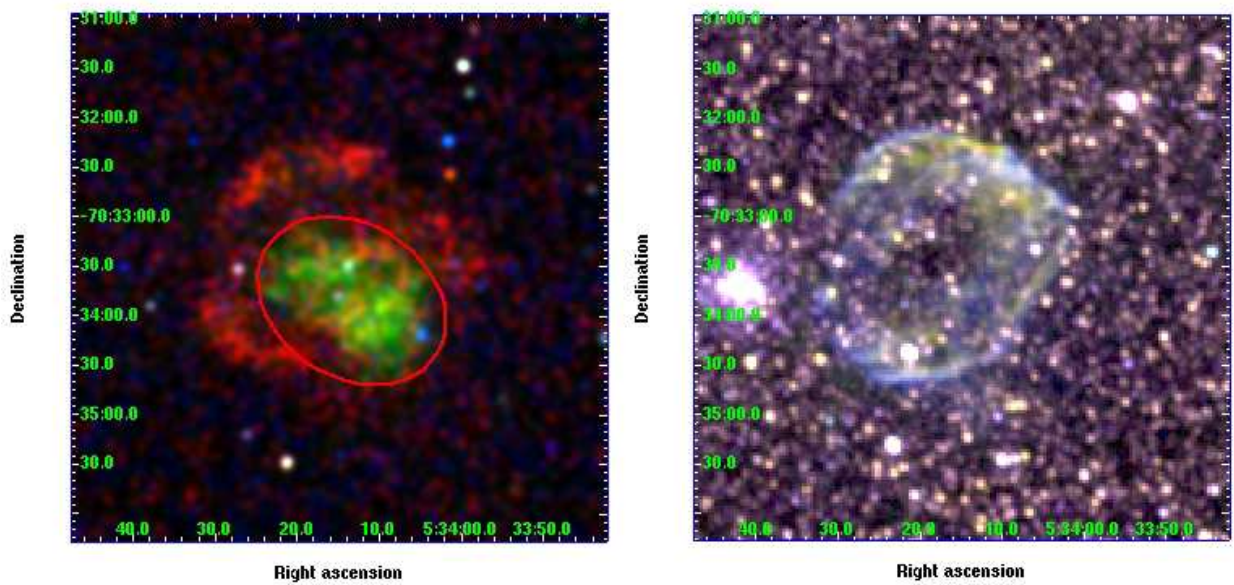}
\caption{
DEM L238 in X-rays ({\it Chandra, left}) and optical (MCELS, 
{\it right}). {\it Chandra} data have been convolved with a
$5\farcs8$ FWHM Gaussian. Soft (0.3--0.7 keV), medium (0.7--3 keV),
and hard (3--7 keV) X-ray photons are coded by red, green, and
blue, respectively. MCELS [\ion{S}{2}] $\lambda\lambda 6716,6731$, H$\alpha$, 
and [\ion{O}{3}] $\lambda 5007$ images are shown in red, green, and blue,
respectively.
The central spectral extraction region is shown in the {\it Chandra}
X-ray image.
\label{fig2}}
\end{figure}

\begin{figure}
\epsscale{1.0}
\plotone{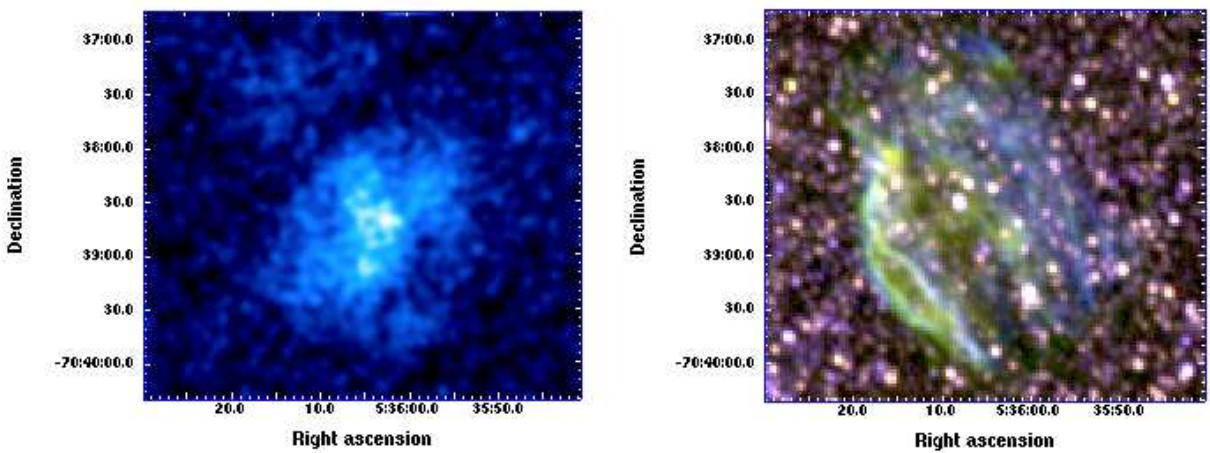}
\caption{
DEM L249 in X-rays ({\it Chandra, left}) and optical (MCELS, {\it
right}). The 0.3--5 keV {\it Chandra} image has been convolved with a
$5\farcs8$ FWHM Gaussian. MCELS [\ion{S}{2}] $\lambda\lambda 6716,6731$, H$\alpha$, 
and [\ion{O}{3}] $\lambda 5007$ images are shown in red, green, and blue, 
respectively.
\label{fig2a}}
\end{figure}

\begin{figure}
\epsscale{.80}
\plotone{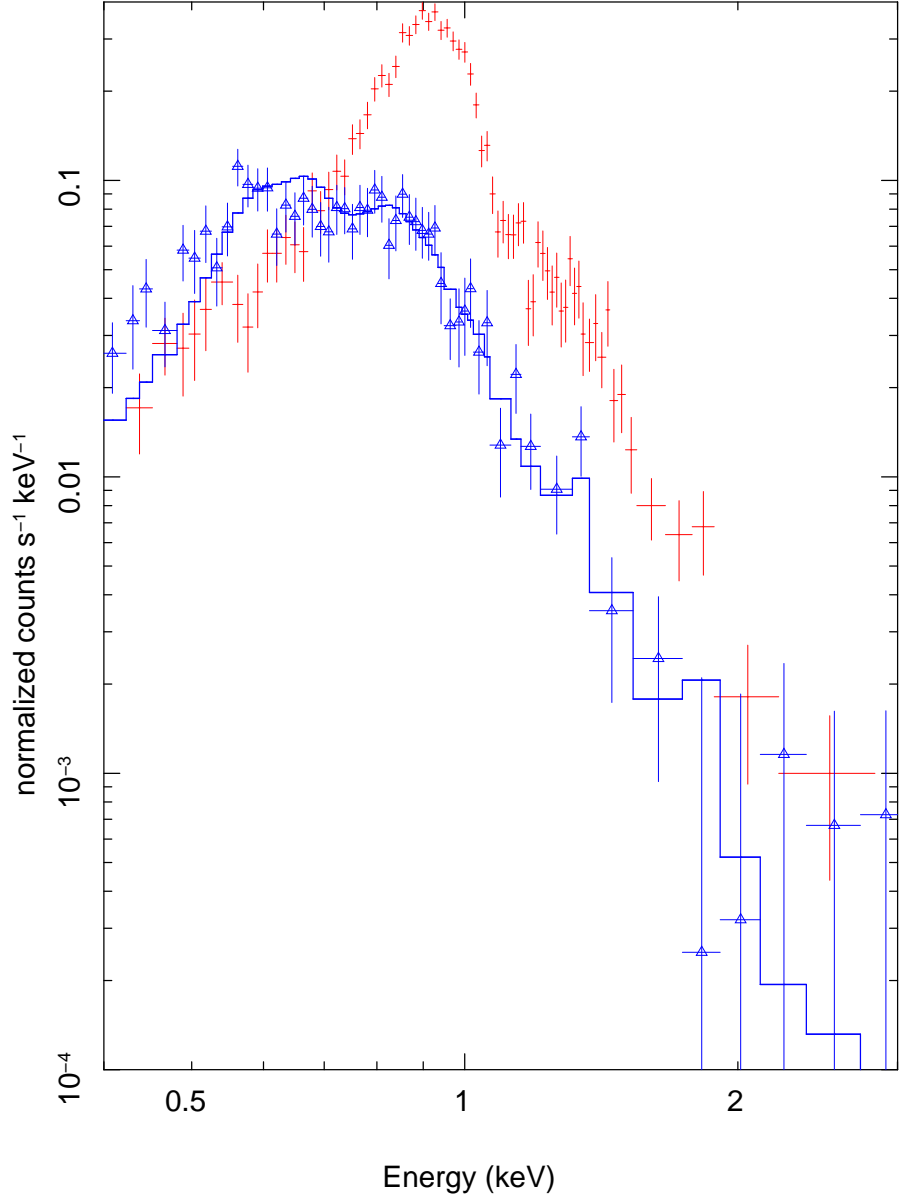}
\caption{Comparison of the central ({\it plus signs}) and shell ({\it
triangles}) {\it Chandra} spectra of DEM L238.  Prominent Fe L-shell
emission dominates the central spectrum.  The shell spectrum is fit
with a Sedov model. \label{fig3}}
\end{figure}

\begin{figure}
\epsscale{.80}
\plotone{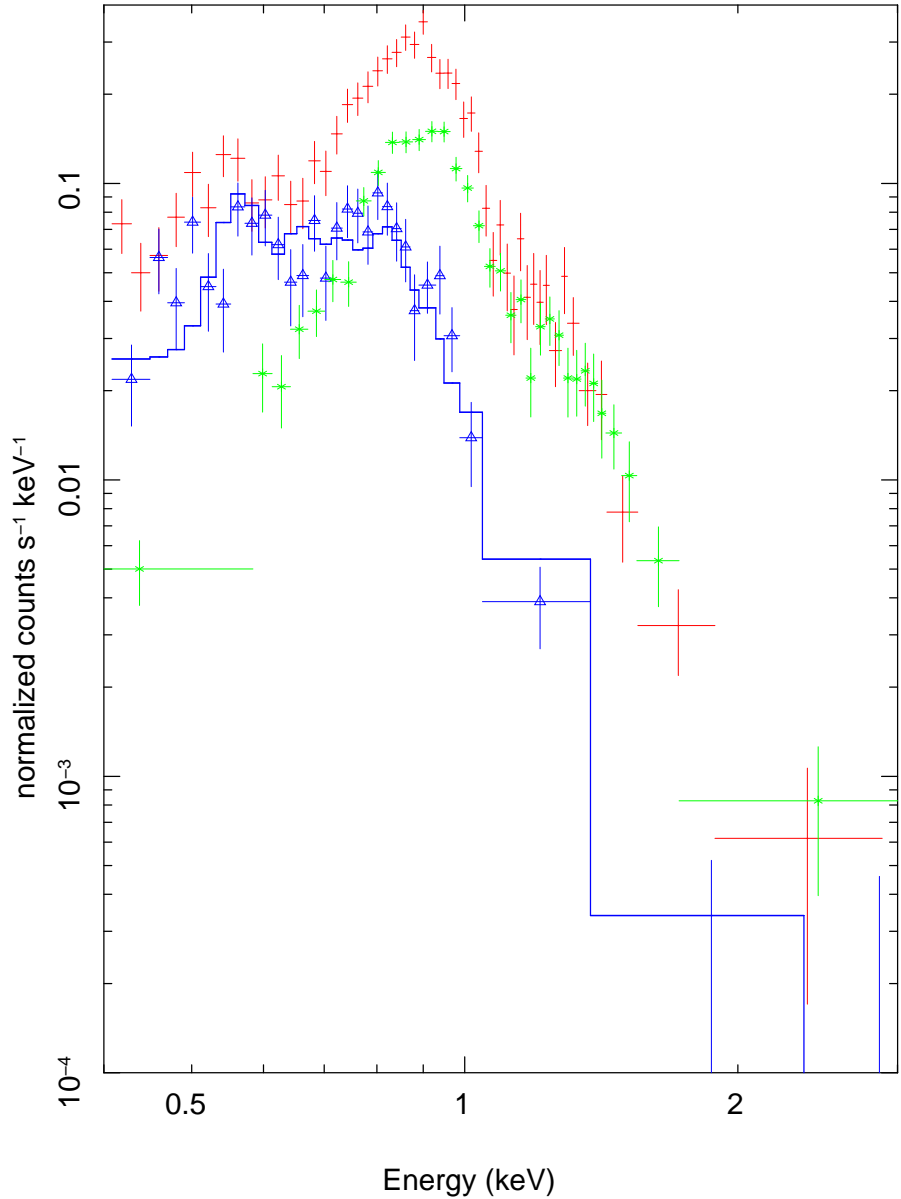}
\caption{Comparison of the central ({\it plus signs}) and shell ({\it
triangles}) {\it XMM-Newton} EPIC pn spectra of DEM L249. A prominent
Fe L-shell emission dominates the central spectrum.  The shell
spectrum is fit with a Sedov model. The spatially integrated {\it
Chandra} spectrum ({\it stars}) is also shown. \label{fig4}}
\end{figure}

\begin{figure}
\epsscale{.80}
\plotone{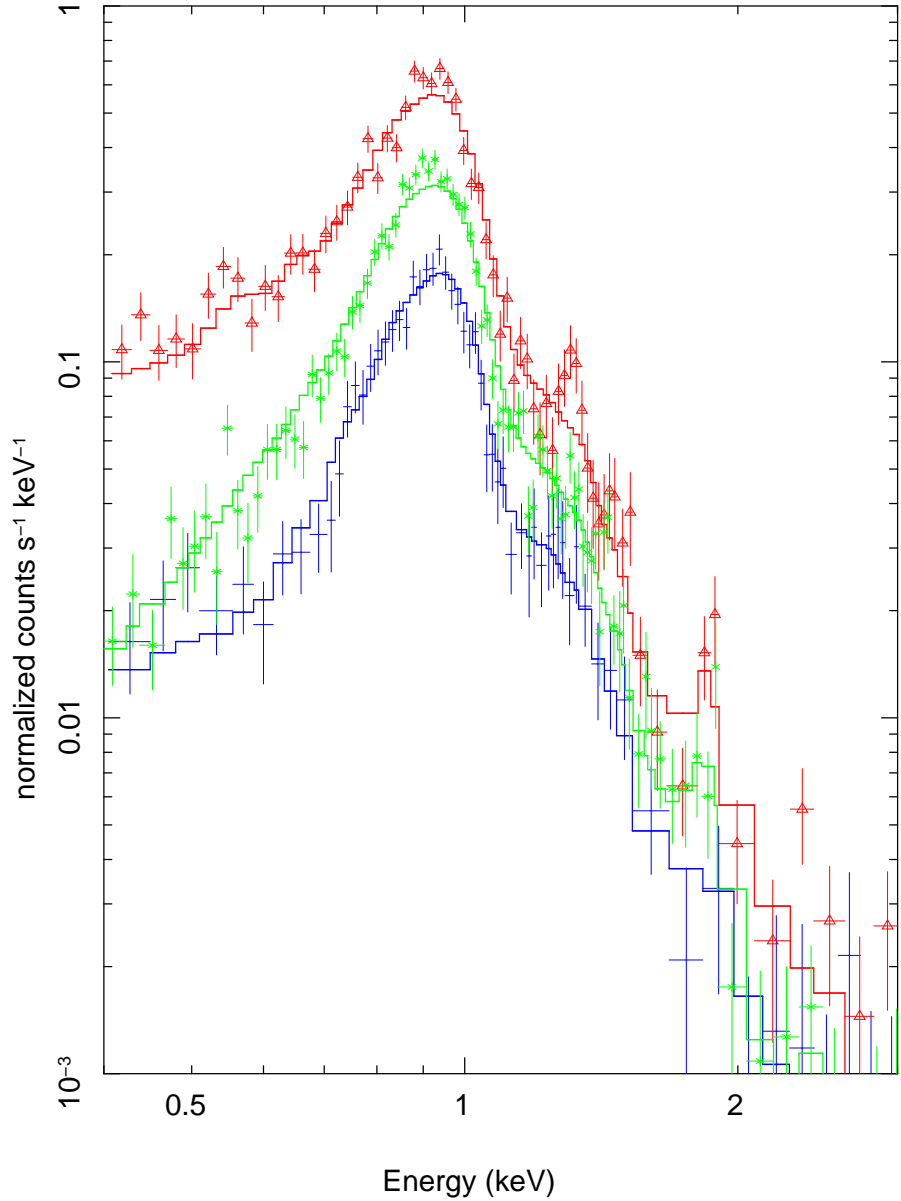}
\caption{Central spectra of DEM L238 ({\it XMM-Newton} pn spectrum, {\it top};
combined
MOS spectra, {\it bottom}; {\it Chandra} spectrum,
{\it middle}). All three model fits require Fe abundance in excess
of solar with respect to other elements. \label{fig5}}
\end{figure}

\begin{figure}
\epsscale{.80}
\plotone{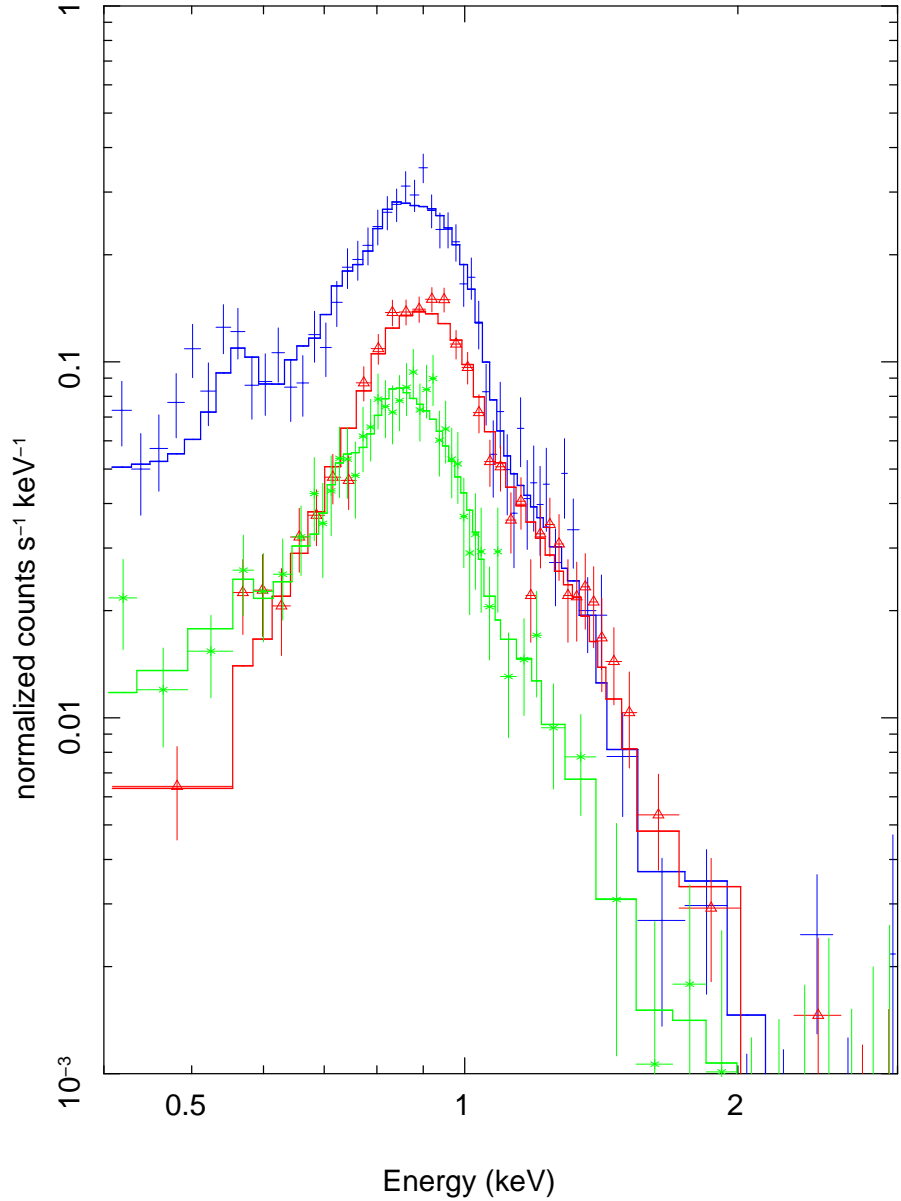}
\caption{Central spectra of DEM L249 ({\it XMM-Newton} pn and combined
MOS spectra, {\it top} and {\it bottom}; {\it Chandra} spectrum,
{\it middle}: triangles). All three model fits require an Fe abundance in excess
of solar with respect to other elements. \label{fig6}}
\end{figure}

\begin{figure}
\epsscale{.80}
\plotone{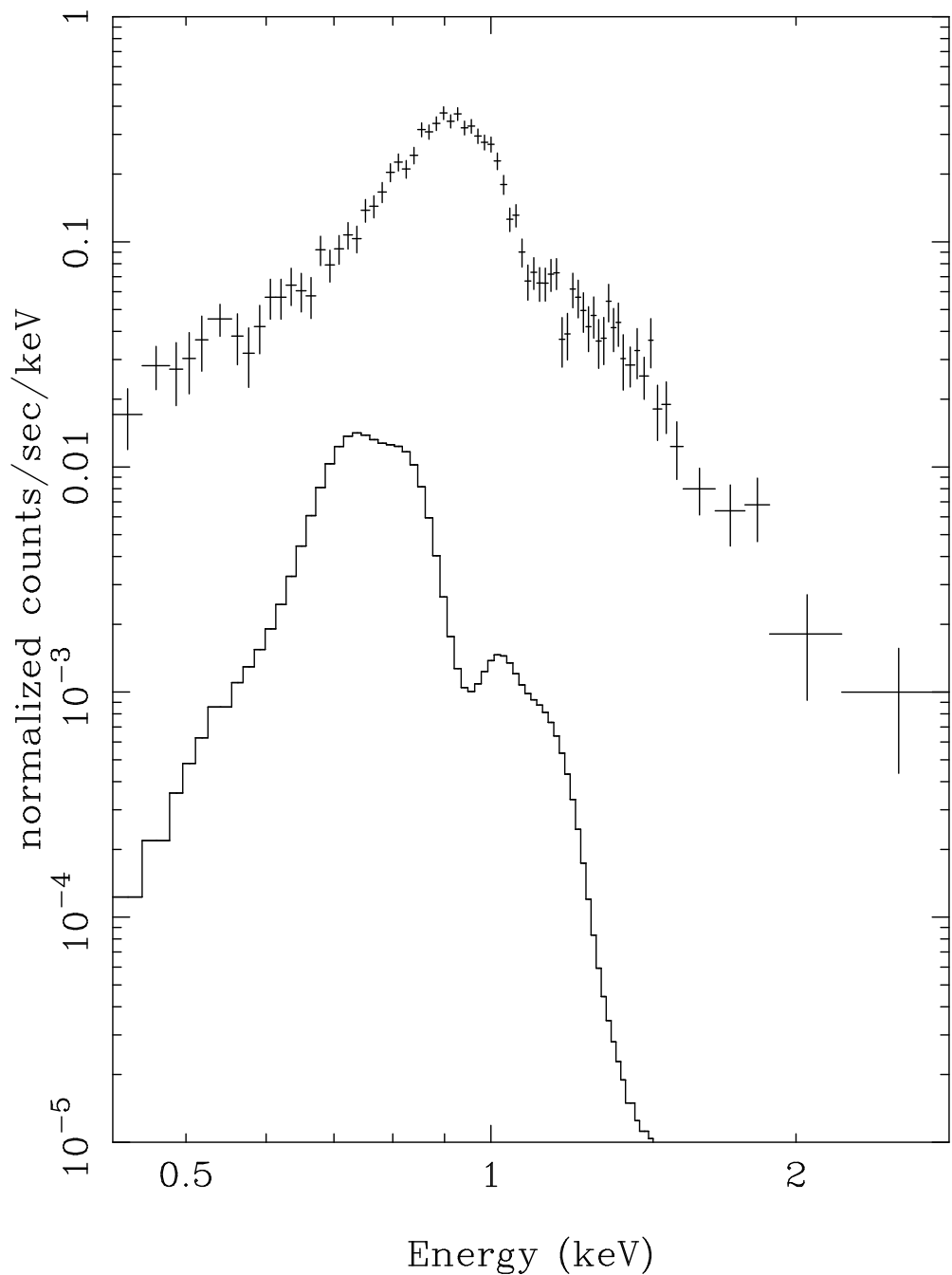}
\caption{Central {\it Chandra} spectrum of DEM L238 ({\it
plus signs}) compared with a model spectrum of ejecta ({\it solid line})
based on the simple Type Ia SNR model discussed in the text. The ejecta
consist of 1.4 M$_\odot$ of pure Fe. A dramatic discrepancy with model
predictions is apparent.
\label{fig7}}
\end{figure}

The presence of denser than expected gas
at the center of DEM L238 and DEM L249 might be explained if they resulted
instead from CC explosions.  Dense
circumstellar medium (CSM) has been found close to progenitors of CC
explosions, including SN 1987A.  The CSM will be shocked early in a
SNR's evolution, and at late times the shocked CSM gas will be near 
CIE conditions, or maybe even recombining \citep{im89}. After the passage
through the CSM, the blast wave will enter a bubble blown by the SN progenitor 
during its main-sequence stage, and then it will hit a dense swept-up ISM 
shell surrounding the tenuous bubble. This could explain the presence of dense, 
optically-emitting material in Figures \ref{fig2} and \ref{fig2a}. The
problem with this hypothesis is that Fe is not the dominant nuclear
burning product of CC explosions. The observed Fe overabundance would
require the operation of a selective mechanism that would enhance
emission from Fe and suppress emission from abundant elements such as
O. This is not likely, and we reject a CC origin for DEM L238 and
DEM L249. Another possibility involves an increased Fe abundance in
the LMC gas itself, but there is no evidence for such large Fe
overabundances in the LMC.

Perhaps Fe is produced in thermonuclear explosions, but the
ejecta structure, the structure of the ambient medium, or both are quite
different than in the standard Type Ia models
\citep{dwarkadas98,badenes03,badenes05,badenes06}.  Type Ia progenitors 
are still poorly
understood, and there is growing evidence for the existence of peculiar
Type Ia SNe.  A subluminous class of Type Ia SNe was recently
discovered \citep{jha06}, exemplified by SN 2002cx, with low expansion
velocities and the presence of relatively dense partially burned or
unburned material at their centers.  Dense, slow-moving Fe-rich gas is
likely to be prominent in the interiors of remnants of such explosions,
although detailed comparisons with observations of DEM L238 and DEM
L249 must await a better understanding of this peculiar class of
SNe. All five known members of the class of SN 2002cx exploded in blue,
star-forming spiral galaxies, making them candidates for the prompt
explosions suggested by \citet{mdp06} and \citet{sullivan06}. This and 
perhaps other classes
of prompt Type Ia SNe must originate from more massive progenitors
than standard, much older Type Ia SNe. Such progenitors are more
likely to be surrounded by circumstellar material formed by ejection
of the massive progenitor envelope prior to the explosion. Several
Type Ia SNe have been found to be surrounded by dense 
CSM, including SN 2002ic \citep{hamuy03}, although such SNe are
infrequent. Among remnants of historical supernovae, Kepler's SNR
shows evidence for dense circumstellar material
\citep[e.g.,][]{blair05}, and \citet{kinugasa99} suggested a Type Ia
origin for this remnant based on a high Fe abundance derived from its
{\it ASCA} X-ray spectrum.  Finally, if the companion to a white dwarf
was a massive asymptotic giant branch star at the time of the white
dwarf explosion, its entire envelope could have been stripped by SN
ejecta \citep{marietta00}. At late stages of SNR evolution, an
Fe-contaminated envelope is expected to appear as dense Fe-rich gas
close to the center of the SNR.  An asymmetric morphology is expected
in this case.

\citet{chuken88} associated DEM L238 and DEM L249 with old (Population
II) stars and Type Ia explosions, because a CC origin for these SNRs is 
unlikely based on their large projected distances from OB associations 
in the LMC.  But $\sim 10^8$ yr old Type Ia progenitors belonging to the
young stellar population would be too old to be associated with the much 
younger ($\la 10^7$ yr) massive stars in OB associations. Cepheid variables
are better tracers of stars with such ages, and their spatial distribution
is not correlated with OB associations and is only very marginally correlated
with older star clusters \citep{batefr99}. DEM L238 and DEM L249 are located 
in the LMC bar with a high Cepheid density \citep{alcock99}. It is possible 
that their progenitors were born in the same star formation episode as 
Cepheids, although older progenitors belonging to Population I (or even to 
Population II) cannot be excluded, as this region of the LMC consists of 
a mixture of stellar populations.

DEM L238 and DEM L249 might belong to the same SNR population as young
Fe-rich SNRs N103B and Kepler. The circumstellar gas in
optically emitting N-rich radiative shocks in Kepler is dense;
\citet{blair91} estimated preshock densities in the 100 -- 1000
cm$^{-3}$ range. This gas traces the densest parts of the clumpy
circumstellar material, where radiative cooling is important. The
primary non-radiative blast wave propagates through the bulk of the
strongly asymmetric CSM, with preshock densities as high as $\sim 10$
cm$^{-3}$ \citep{blair91}. Optically emitting radiative shocks with
high (180 cm$^{-3}$) preshock densities are also present in N103B
\citep{russell90}. Unlike in Kepler, the N abundance is similar to its
ISM value, but densities are much higher than in the surrounding \ion{H}{2}
region N103 or in the general diffuse ISM. Such high densities are more
typical of molecular clouds, but there are no known nearby molecular
clouds associated with N103B \citep{fukui99,mizuno01}. These high densities 
and the high X-ray luminosity of N103B suggest the presence of dense 
circumstellar gas, although an ISM origin for the ambient gas cannot be
ruled out at this time. N103B may be associated with the nearby double star
cluster NGC 1850. The main cluster is 50 Myr old, but the second cluster is 
very young \citep[2--5 Myr;][]{calcas98}. Optical observations 
revealed an enhancement in the N103B H$\alpha$ emission at 
the radial velocity of the \ion{H}{2} region N103 produced by young stars in NGC 1850 
and its surroundings \citep{chuken88}. \citet{chuken88} interpreted this as 
evidence for a
blast wave interaction with ambient ISM gas, but an alternative 
explanation involves the CSM around a moderately old (50 Myr) SN progenitor 
with a radial velocity comparable to its parent cluster (and hence the \ion{H}{2} 
region N103). Just like Kepler, N103B is strongly asymmetric at all
wavelengths, with a compact and bright western hemisphere and a large
and much fainter eastern hemisphere. The asymmetry in the ambient CSM
has been invoked to explain the asymmetry of the Kepler SNR, and a
similar explanation is likely for N103B. At an age of $\sim 10^4$ yr,
both Kepler and N103B could still retain an imprint of the dense
ambient CSM, and resemble DEM L238 and DEM L249.

In summary, {\it Chandra} and {\it XMM-Newton} observations revealed
Fe-rich gas at the centers of LMC SNRs DEM L238 and DEM L249. While
the Fe overabundance suggests Type Ia progenitors, the standard Type
Ia models cannot explain the presence of relatively dense SN ejecta
with long ionization timescales.  At this time we do not understand
the origin of these ejecta; one possibility involves prompt Type
Ia explosions with progenitors more massive than average. The remnant
of Kepler's SN and the young LMC SNR N103B may belong to this
category of Type Ia SNRs.  Studies of remnants of such explosions in
the MCs promise to shed light on the nature of their poorly understood
progenitors. The presence of such unusual progenitors may also force a
reevaluation of the current chemical evolution models of the MCs, in
view of the prominent role of Type Ia explosions in their evolution.

\acknowledgments We thank John Blondin for his VH-1 hydrocode, and
Paul Plucinsky for his help with planning {\it Chandra}
observations. Optical images are courtesy of Sean Points, Chris Smith, and
the Magellanic Cloud Emission Line Survey (MCELS) collaboration. This work 
was supported by NASA through grants NAG5-13641, SAO G03-4097X, and 
SAO AR5-6007X.

\clearpage



\clearpage

\clearpage

\clearpage

\clearpage

\clearpage

\clearpage

\def\res#1#2#3{$#1_{-#2}^{+#3}$}
\begin{deluxetable}{lcccccc}

\tablecolumns{7}

\tablewidth{0pc}

\tabletypesize{\footnotesize}

\tablecaption{Fits to the Central Region of DEM L238\label{tabl238}}

\tablehead{

  \colhead{}  & \multicolumn{3}{c}{{\tt vpshock}} & \multicolumn{3}{c}{{\tt Sedov} $+$ {\tt vapec}}  \\

  \colhead{Parameters} & ACIS & pn & MOS & ACIS & pn & MOS }

\startdata
$\chi^2$/DOF & 73/83 & 148/112 & 72/70 & 87/83 & 156/112 & 77/70 \\

$(EM_S/4\pi d^2)/10^{10}$ cm$^{-5}$ & - & - & - & \res{0.45}{0.12}{0.12} & \res{0.44}{0.09}{0.09} & \res{0.15}{0.12}{0.14} \\

$kT_e$ keV & \res{0.80}{0.02}{0.03} & \res{0.82}{0.02}{0.04} & \res{0.80}{0.02}{0.03} & \res{0.81}{0.01}{0.01} & \res{0.82}{0.01}{0.01} & \res{0.82}{0.02}{0.02}  \\

Fe (Ni)    & \res{1.25}{0.24}{0.31} & \res{1.15}{0.19}{0.27} & \res{1.28}{0.26}{0.40} & \res{1.51}{0.33}{0.55} & \res{1.29}{0.24}{0.36} & \res{1.39}{0.28}{0.43}  \\

$\tau/10^{12}$ cm$^{-3}$s  & \res{1.17}{0.42}{0.74} & \res{0.83}{0.26}{0.38} & \res{2.3}{1.2}{7.1} & - & - & - \\
$(EM_C/4\pi d^2)/10^{10}$ cm$^{-5}$ & 1.81  & 1.38 & 1.40 & \res{1.52}{0.39}{0.40} & \res{1.24}{0.26}{0.26} & \res{1.31}{0.30}{0.30} \\

\enddata

\end{deluxetable}


\def\res#1#2#3{$#1_{-#2}^{+#3}$}
\begin{deluxetable}{lcccccc}

\tablecolumns{7}

\tablewidth{0pc}

\tabletypesize{\footnotesize}

\tablecaption{Fits to the Central Region of DEM L249\label{tabl249}}

\tablehead{

  \colhead{}  & \multicolumn{3}{c}{{\tt vpshock}} & \multicolumn{3}{c}{{\tt Sedov} $+$ {\tt vapec}}  \\

  \colhead{Parameters} & ACIS & pn & MOS & ACIS & pn & MOS }

\startdata
$\chi^2$/DOF & 52/79 & 71/65 & 32/42 & 54/79 & 63/65 & 32/42 \\

$(EM_S/4\pi d^2)/10^{10}$ cm$^{-5}$ & - & - & - & \res{0.59}{0.21}{0.21} & \res{0.32}{0.06}{0.06} & \res{0.34}{0.10}{0.10} \\

$kT_e$ keV & \res{0.75}{0.06}{0.18} & \res{1.25}{0.50}{0.17} & \res{0.71}{0.05}{0.25} & \res{0.73}{0.04}{0.04} & \res{0.77}{0.02}{0.02} & \res{0.69}{0.02}{0.03}  \\

Fe (Ni)    & \res{1.27}{0.37}{0.71} & \res{3.5}{2.3}{0.8} & \res{1.37}{0.44}{1.50} & \res{1.17}{0.36}{0.78} & \res{1.79}{0.55}{1.28} & \res{1.17}{0.39}{0.95}  \\

$\tau/10^{11}$ cm$^{-3}$s  & \res{3.7}{2.2}{4.5} & \res{0.88}{0.20}{5.30} & \res{4.2}{3.1}{4.1} & - & - & - \\
$(EM_C/4\pi d^2)/10^{10}$ cm$^{-5}$ & 1.62 & 0.34 & 1.03 & \res{1.59}{0.63}{0.67} & \res{0.82}{0.34}{0.34} & \res{1.06}{0.46}{0.42} \\

\enddata

\end{deluxetable}

\end{document}